\documentclass[graybox]{svmult}

\usepackage{mathptmx}       
\usepackage{helvet}         
\usepackage{courier}        
\usepackage{type1cm}        
\usepackage{makeidx}         
\usepackage{graphicx}        
\usepackage{multicol}        
\usepackage[bottom]{footmisc}

\makeindex             

\begin{document}
\newcommand\mnras{MNRAS}
\newcommand\pasp {PASP}
\newcommand\aj  {AJ}
\newcommand\apj {ApJ}
\newcommand\apjl {ApJL}
\newcommand\apjs {ApJS}
\newcommand\aap {A\&A}
\newcommand\aaps {A\&AS}
\newcommand\apss {AP\&SS}
\newcommand\araa {ARA\&A}
\newcommand\ppv {PPV}
\newcommand\ppII {PPII}

\title*{Do All Stars in the Solar Neighbourhood Form in Clusters?}
\titlerunning{Do All Stars Form in Clusters?} 
\author{Eli Bressert, Nate Bastian \& Robert Gutermuth}
\institute{Eli Bressert \at University of Exeter, School of Physics, Stocker Road, Exeter EX4 4QL, \email{eli@astro.ex.ac.uk}
\and Nate Bastian \at University of Exeter, School of Physics, Stocker Road, Exeter EX4 4QL \email{bastian@astro.ex.ac.uk}
\and Robert Gutermuth \at Univerity of Massachusetts, Smith College, Northampton, MA 01063, USA \email{rgutermu@smith.edu}
}


%
%
\maketitle

\vskip-1.2truein

\abstract{We present a global study of low mass, young stellar object (YSO) surface densities ($\Sigma$) in nearby ($< 500$~pc) star forming regions based on a comprehensive collection of {\it Spitzer Space Telescope} surveys. We show that the distribution of YSO surface densities is a smooth distribution, being adequately described by a lognormal function from a few to $10^3$ YSOs per $\textrm{pc}^{2}$, with a peak at $\sim22$~stars $\textrm{pc}^{-2}$. The observed lognormal $\Sigma$ is consistent with predictions of hierarchically structured star-formation at scales below 10~pc, arising from the molecular cloud structures.  We do not find evidence for multiple discrete modes of star-formation (e.g.~clustered and distributed). Comparing the observed $\Sigma$ distribution to previous $\Sigma$ threshold definitions of clusters show that they are arbitrary. We find that only a low fraction ($< 26$\%) of stars are formed in dense environments where their formation/evolution (along with their circumstellar disks and/or planets) may be affected by the close proximity of their low-mass neighbours.}

\section{Introduction}
\label{sec:1}

It is often stated that most if not all stars form in clusters. This view is based largely on near-infrared (NIR) studies of star-forming (SF) regions within several hundred parsecs of the Sun \cite{Lada2003,Porras2003}. However, combining high-resolution mid-infrared (MIR) data with the NIR makes YSO identification more robust and less likely to suffer from field star contamination, which leads to better tracing of YSO surface densities. This means that with the NIR alone, there were large uncertainties in the number of stars at low values of YSO surface densities ($\Sigma_{\rm YSO}$) \cite{Carpenter2000}.

With the launch of the {\it Spitzer Space Telescope} \cite{Werner2004} we are now able to differentiate YSOs and contaminating sources based on colour information and hence can study the distribution of YSOs independently of the surface densities. Large field-of-view (FoV) {\it Spitzer} observations of SF regions \cite{Allen2007,Evans2009} found that YSOs extend well beyond the densest groups in their environment and continue throughout. Several {\it Spitzer} surveys that cover nearly all the SF regions within 500~pc of the Sun were combined and presented by \cite{Bressert2010}.

The spatial distribution of forming stars, i.e. do they form in clusters, is important for many reasons and here we list several. What spatial distributions of star forming environments lead the Galactic stellar clusters and why do they constitute such a low fraction of the stellar population? Are there multiple discrete modes of star formation, such as clustered and distributed, that manifest themselves as peaks in a surface density distribution (e.g. \cite{Strom1993,Carpenter2000,Weidner2004,Wang2009})? What fraction of star forming environments are dense enough to affect the YSOs and alter their disk and planet formation/evolution \cite{Allen2007,Gutermuth2005}, which we refer to as \emph{dense environments}.

Outside the Milky Way we see that star-formation proceeds hierarchically from $\sim10$~pc (the resolution limit of most extragalactic surveys) to kpc scales (e.g. \cite{Efremov1995, Elmegreen1996, Bastian2007, Bastian2009, Gieles2008}). If star formation occurs hierarchically down to 0.1~pc, then it is unlikely, and indeed not expected, that all stars form in gravitationally bound young stellar clusters.  The availability of large FoV {\it Spitzer} surveys of nearby star-forming regions allows us to test how far down in scale this hierarchy proceeds. By hierarchical structure we mean a smoothly varying non-uniform distribution of densities, where denser subareas are nested within larger, less dense areas \cite{Scalo1985,Elmegreen2006}. 

In this paper we will investigate 1$)$ whether there is evidence for multi-modality in the surface densities of YSOs, 2$)$ what fraction of stars form in dense environments in the local neighbourhood and 3$)$ how relevant the various cluster definitions are as provided by \cite{Carpenter2000,Lada2003,Allen2007,Jorgensen2008,Gutermuth2009}.

\begin{figure}
\includegraphics[width=8cm]{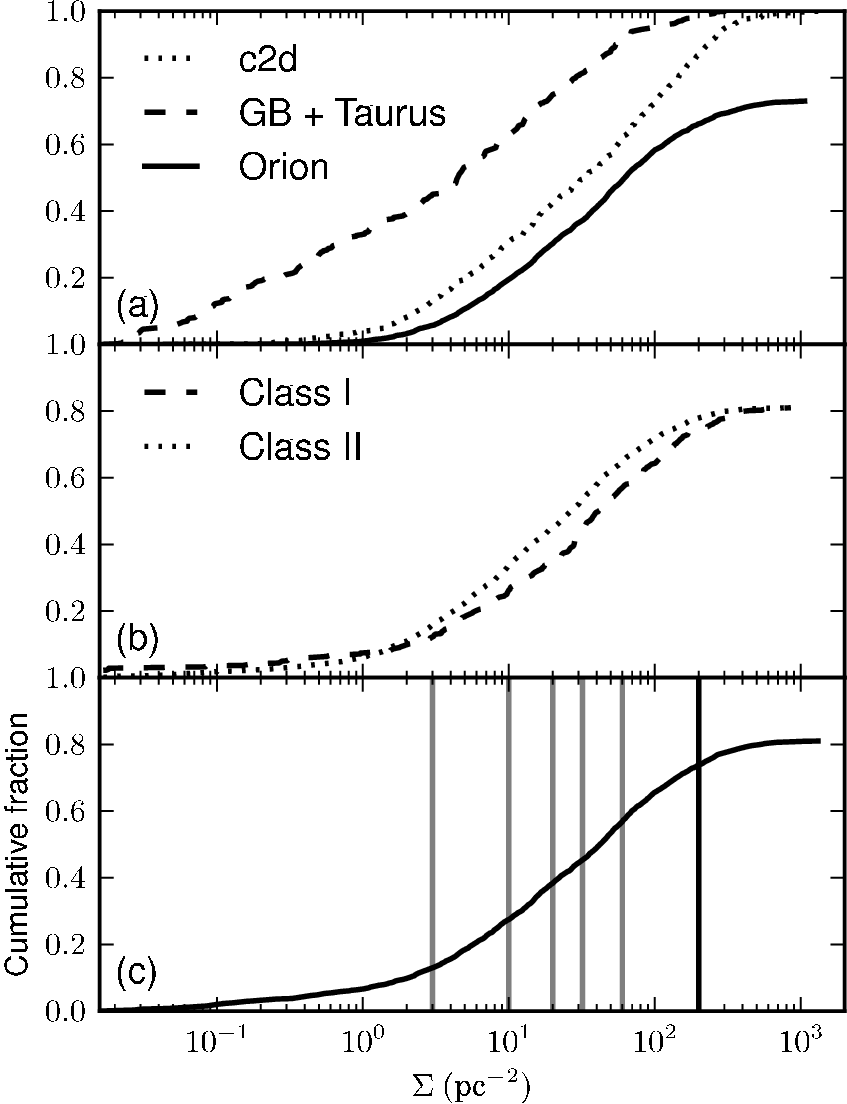} 
\caption{{\bf(a)} The cumulative fraction of surface densities for the GB+Taurus, c2d, and Orion surveys. Each SF region included in the distributions has N(YSOs) $\geq 10$ and a sufficient field-of-view to properly calculate stellar surface densities. The Orion survey stops at 73\% for the cumulative fraction since the ONC is excluded. We adopt a 65\% disk fraction for all of the SF regions.  We normalised each curve by the number of YSOs in each survey. {\bf(b)} With the GB+Taurus, c2d, and Orion surveys combined we see Class I \& II distributions having similar profiles with a small offset in density, showing that we are likely seeing the primordial distribution of the YSOs. {\bf(c)} With all of the {\it Spitzer} surveys combined we compare several cluster definitions. The vertical grey lines from left to right are Lada \& Lada (2003), Megeath et al. (in prep.), J{\o}rgensen et al. (2008), Carpenter (2000), and Gutermuth et al. (2009) stellar density requirements for clusters. These values correspond to 3, 10, 20, 32, and 60 YSOs $\textrm{pc}^{-2}$ and intersect the corrected cumulative distribution profile, implying that 87\%, 73\%, 62\%, 55\%, and 43\% of stars form in clusters, respectively. The percentages correlate to what fraction of stars form in `clusters' based on the various definitions. The black vertical line is for a dense environment where $\Sigma \geq$ 200 YSOs$/\textrm{pc}^{2}$. The fraction of YSOs in a dense environment is $<26$\%.} \label{fig:1}
\end{figure}

\section{Data \& Biases} \label{sec:2}

Multiple {\it Spitzer} surveys were used to generate a comprehensive and statistically significant dataset to investigate the spatial surface density properties of forming stars in the solar neighbourhood. The surveys are the Gould's Belt (GB) survey (Allen et al. in prep.), Orion survey (Megeath et al. in prep.), Cores to Disks (c2d) survey \cite{Evans2003}, and the Taurus survey \cite{Rebull2010}. The GB and Orion catalogs have not been publicly released yet. There are more than 7000 YSO detections in the combined catalogs at distances between 100 to 500~pc.

{\it Spitzer} data are necessary for this study as low $\Sigma_{\rm YSO}$ can be differentiated from field star populations, unlike NIR observations where field star contamination can be problematic. The comprehensive YSO population represents a global view of the low-mass star-forming region in the local neighbourhood from low to high surface densities. These {\it Spitzer} surveys combined represent the most complete census of star formation within 500~pc of the Sun available to date.

Data treatment on distances, exclusion of the central region of the Orion Nebular Cluster, rejection of Class III objects, and potential contaminants are described in \cite{Bressert2010}. After the data correction process, we have 3857 YSOs remaining to estimate the surface densities presented in this paper.

\section{$\Sigma_{\rm YSO}$ distributions} \label{sec:3}
Our primary tool for analysing the surface densities is computing the local observed surface density of YSOs centred on each YSO's position, where $\Sigma_{\rm YSO} = (\textrm{N}-1)/(\pi D_{N}^2)$ and N is the Nth nearest neighbour, and $D_{N}$ is the projected distance to that neighbour (see \cite{Casertano1985}). Throughout this work we will adopt $\textrm{N}=7$, although we note that all results have been tested for $\textrm{N}=4-22$ and no significant differences were found. Figure~\ref{fig:1} shows the surface density distribution of all YSOs in our sample.

In order to see the fraction of YSOs above a given $\Sigma$ threshold, we show the combined $\Sigma_{\rm YSO}$ distribution (shown as a cumulative fraction normalised to the number in each combined survey) for the three surveys used in this study in Fig.~\ref{fig:1}a. Note that the GB/Taurus distribution lies to the left of the c2d survey. This is simply due to the GB/Taurus focussing on lower density regions than c2d. The cumulative distribution for the Orion survey only reaches 0.73 in Fig.~\ref{fig:1}a and 0.81 in Fig.~\ref{fig:1}c, where all the surveys have been combined, due to the exclusion of the ONC. In Fig.~\ref{fig:1}c we show the cumulative distribution of all YSOs included in our survey, while in Fig.~\ref{fig:1}b we split the survey into class I and class II objects.

\begin{figure}
    \includegraphics[width=8cm]{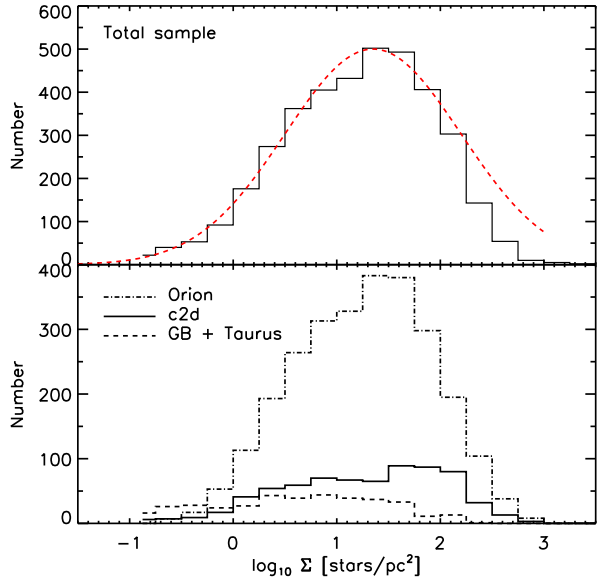} \caption{{\bf Top panel:} The surface density distribution of the total sample of YSOs in the solar neighbourhood used in this work (black).  A lognormal function with a peak at $\sim22$~YSOs/\textrm pc$^2$ and a dispersion $\sigma_{{\rm log}_{10} \Sigma} = 0.85$ is shown as a dashed line. {\bf Bottom panel:} The same as the top panel but now broken into the three respective surveys.}
    \label{fig:2} 
\end{figure}

\section{Results} \label{sec:4} 

It has been long assumed that two distinct modes of star formation exist for YSOs, `clustered' and `distributed' (e.g. \cite{Gomez1993,Carpenter2000,Lada2003}), but the notion has been questioned after \emph{Spitzer} results hinted otherwise \cite{Allen2007}. If there are indeed two modes, then we would expect to see a bi- or multi-modal profile in cumulative surface density distribution plots such as Figs. ~\ref{fig:1}a ands \& ~\ref{fig:1}c. Instead we see smooth and featureless distributions from the low to high stellar surface densities for the c2d, GB, Taurus, and Orion surveys. We find that the $\Sigma_{\rm YSO}$ distribution of low-mass stars in the solar neighbourhood can be well described by a lognormal function, as seen in \ref{fig:2}, with a peak at $\sim22$~YSOs/\textrm pc$^2$ and a dispersion $\sigma_{{\rm log}_{10} \Sigma} = 0.85$.

The spatial distribution of the YSOs in these SF regions is expected to be close to primordial since their YSOs, in particular Class Is and Class IIs, are $\le 2$ Myr old \cite{Haisch2001,Hernandez2007}. In order to place stricter constraints on this, we now split the complete sample into Class I and II objects, which can be roughly attributed to an age sequence. The cumulative $\Sigma$ distributions of Class I and II YSOs are shown in Fig.~\ref{fig:1}b. We see that the two distributions have similar smooth density spectra, however they are slightly offset. The $\Sigma$ of the Class I/II objects are calculated by finding a YSO's Nth nearest YSO. Once this is done for the YSOs we separate the Class I/II objects. $\Sigma$ is calculated this way since Class Is and Class IIs  are not always spatially distinct from one another \cite{Gutermuth2009}. Class IIs are known to be slightly more dispersed than Class Is in high density regions \cite{Gutermuth2009} reflecting early dynamical evolution. However, the similar distribution between these classes leads us to conclude that the distribution of observed $\Sigma$ is mainly primordial in nature.

In Fig.~\ref{fig:1}c we show five vertical grey lines that refer to the defined densities required for a collection of YSOs to be considered `clustered' (Lada \& Lada~2003; Megeath et al. in prep.; J{\o}rgensen et al.~2008; Carpenter~2000, Gutermuth et al.~2009). The vertical lines fall on the same featureless slope and do not correspond to any preferred density. The black vertical line, which corresponds to dense environments (as defined in \cite{Gutermuth2005}), shows that $<26$\% of YSOs are formed in environments where they (along with their disks and planets) are likely to interact with their neighbours.

\section{Discussion and Conclusions} \label{sec:5}

We conclude that stars form in a broad and smooth spectrum of surface densities and do not find evidence for discrete modes of star formation in the $\Sigma$ of low mass YSOs forming in the solar neighbourhood. The observed lognormal surface density distribution is consistent with predictions of hierarchically structured star-formation, which arise from similar structures in molecular clouds. (\cite{Elmegreen2002,Elmegreen2008}). Our results suggest that clusters are not fundamental units in the star formation process, but simply the high density tail end of a continuous distribution. A small fraction ($< 26$\%) of stars form in dense environments where their formation and/or evolution is expected to be influenced by their surroundings.

By comparing our global surface density distribution with the summary of cluster definitions in \cite{Bressert2010}, we find that the fraction of stars forming in clusters is crucially dependent on the adopted definitions (ranging from $\sim40$ to $90$\%). Since star/cluster formation happens in a dynamic environment \cite{Bate2003,Allison2009,Moeckel2010} it may be impossible to define what will become a stellar cluster in these early stages, as the final object that remains bound is simply the dynamically mixed part of a larger, initially hierarchical, distribution. In this scenario a cluster can only be clearly defined above the surrounding distribution once it is dynamically evolved where t$_{\rm age}$/t$_{\rm cross} > 1$, as defined by Gieles \& Portegies Zwart 2010\nocite{Gieles2010}.

Star forming environments provide the initial conditions from which star clusters may eventually form, albeit rarely. Since the probability density function of molecular gas varies with environment, as does the tidal field experienced by the SF regions, it is likely that the fraction of YSOs ending up in bound star clusters varies with environment \cite{Elmegreen2008} and the observed $\Sigma_{\rm YSO}$ is not universal. Future investigations will extend this work out to 2~kpc where proto-stellar cores and high-mass SF regions will be included, and testing whether star formation primarily occurs when $\Sigma_{gas}$ is above a specific threshold (e.g. \cite{Heiderman2010, Andre2010, Johnstone2004})  

\begin{acknowledgement}
We would like to thank the organisers of the JENAM cluster for arranging the meeting and inviting us to contribute our work. Additionally, Eli Bressert's trip to the meeting was possible due to the CONSTELLATION network and funds. 
\end{acknowledgement}


\begin{thebibliography} {99.}%
\bibitem{Allen2007} Allen, L., et al.\ 2007, \ppv
\bibitem{Allison2009} Allison, R.~J., et al.\ 2009, \apjl, 700, L99
\bibitem{Andre2010} Andr\'{e}, P. et al.\ 2010, \aap, 518, L102
\bibitem{Bastian2007} Bastian, N. et al.\ 2007, \mnras, 379, 1302
\bibitem{Bastian2009} Bastian, N. et al.\ 2009, \mnras, 392, 868
\bibitem{Bate2003} Bate, M.~R, Bonnell, I.~A., \& Bromm, V.\ 2003, \mnras, 339, 577
\bibitem{Bressert2010} Bressert, E., et al.\ 2010, \mnras, 409, L54
\bibitem{Carpenter2000} Carpenter, J.~M.\ 2000, \aj, 120, 3139
\bibitem{Casertano1985} Casertano, S. \& Hut, P.\ 1985, \apj, 298, 80
\bibitem{Efremov1995} Efremov, Y.~N.\ 1995, \aj, 110, 2757
\bibitem{Elmegreen1996} Elmegreen, B.~G. \& Efremov, Y.~N.\ 1996, 466, 802
\bibitem{Elmegreen2002} Elmegreen, B.~G.\ 2002, \apj, 564, 773
\bibitem{Elmegreen2006} Elmegreen, B.~G., Elmegreen, D.~M., Chandar, R., Whitmore, B. \& Regan, M.\ 2006, \apj, 644, 879
\bibitem{Elmegreen2008} Elmegreen, B.~G.\ 2008, \apj, 672, 1006
\bibitem{Evans2003} Evans, II, N.~J., et al.\ 2003, \pasp, 115, 965
\bibitem{Evans2009} Evans, II, N.~J., et al.\ 2009, \apjs, 181, 321
\bibitem{Gieles2008} Gieles, M., Bastian, N. \& Ercolano, B.\ 2008, \mnras, 391, L93
\bibitem{Gieles2010} Gieles, M. \& Portegies Zwart, S.~F.\ 2010, \mnras, 410, L6
\bibitem{Gomez1993} Gomez, M., Hartmann, L., Kenyon, S.~J. \& Hewett, R.\ 1993, \aj, 105, 1927
\bibitem{Gutermuth2005} Gutermuth, R.~A., et al.\ 2005, \apj, 632, 397
\bibitem{Gutermuth2009} Gutermuth, R.~A., et al.\ 2005, \apjs, 184, 18
\bibitem{Haisch2001} Haisch, Jr., K.~E., Lada, E.~A. \& Lada, C.~J.\ 2001, \apjl, 553, L153
\bibitem{Heiderman2010} Heiderman, A., Evans, II, N.~J., Allen, L.~E., Huard, T. \& Heyer, M.\ 2010, \apj, 723, 1019
\bibitem{Hernandez2007} {Hern{\'a}ndez}, J., et al.\ 2007, \apj, 671, 1784
\bibitem{Jorgensen2008} J{\o}rgensen, J.~K., Johnstone, D., Kirk, H., Myers, P.~C., Allen, L.~E. \& Shirley, Y.~L.\ 2008, \apj, 683, 822
\bibitem{Johnstone2004} Johnstone, D., Di Francesco, J. \& Kirk, H.\ 2004, \apjl, 611, L45
\bibitem{Lada2003} Lada, C.~J. \& Lada, E.~A.\ 2003, \araa, 41, 57
\bibitem{Moeckel2010} Moeckel, N. \& Bate, M.~R.\ 2010, \mnras, 404, 721
\bibitem{Porras2003} Porras, A., Christopher, M., Allen, L., Di Francesco, J., Megeath, S.~T. \& Myers, P.~C.\ 2003, \aj, 126, 1916
\bibitem{Rebull2010} Rebull L.~M., et al.\ 2010, \apjs, 186, 259
\bibitem{Scalo1985} Scalo, J.~M.\ 1985, \ppII
\bibitem{Strom1993} Strom, K.~M., Strom, S.~E. \& Merrill, K.~M.\ 1993, \apj, 412, 233
\bibitem{Wang2009} Wang, J., Feigelson, E.~D., Townsley, L.~K., Rom{\'a}n-Z{\'u}{\~n}iga, C.~G., {Lada}, E. \& {Garmire}, G.\ 2009, \apj, 696, 47
\bibitem{Weidner2004} Weidner, C., Kroupa, P. \& Larsen, S.~S.\ 2004, \mnras, 350, 1503
\bibitem{Werner2004} Werner, M.~W, et al.\ 2004, \apjs, 154, 1
\end{thebibliography}
\end{document}